\documentclass{ws-procs975x65}

\begin{document}

\title{Quantum phases of supersymmetric lattice models}

\author{L. Huijse$^*$ and K. Schoutens}

\address{Institute for Theoretical Physics, University of Amsterdam,\\
Valckenierstraat 65, 1018 XE Amsterdam, The Netherlands\\
$^*$E-mail: L.Huijse@uva.nl}

\begin{abstract}
We review recent results on lattice models for spin-less fermions with strong repulsive
interactions. A judicious tuning of kinetic and interaction terms leads to a model
possessing supersymmetry. In the 1D case, this model displays critical behavior
described by superconformal field theory. On 2D lattices we generically find
superfrustration, characterized by an extensive ground state entropy. For certain 2D
lattices analytical results on the ground state structure reveal yet another quantum
phase, which we tentatively call 'supertopological'.
\end{abstract}


\bodymatter

\section{Introduction}\label{sec1}

A long standing challenge in condensed matter physics is to understand the
properties of strongly correlated electron systems. While it is relatively easy
to formulate model descriptions, it has proved exceedingly difficult to arrive at
exact results for these in spatial dimensions $D>1$, in particular in regimes where
interaction and kinetic terms in the relevant hamiltonian are of comparable strength.
In a series of papers, initiated by one of the authors together with P.~Fendley
and J.~de Boer \cite{fendley-2003-90},
a specific model for lattice fermions was explored, precisely
in this non-perturbative regime. The key property of this model is {\bf supersymmetry}.
This gives a subtle balance between kinetic and interaction terms, leading to
remarkable features such as, in particular, large degeneracies of quantum ground states.
At the same time, supersymmetry provides a rich mathematical structure that can be
employed to derive rigorous results for some of the key features of the model.

The degrees of freedom are spin-less fermionic particles living on
a given lattice. A fermion at site $i$ is created by the operator
$c_i^\dagger$ with $\{c_i,c^\dagger_j\}=\delta_{ij}$. The fermions
have a {\it hard core}, meaning that the presence of a fermion on
a given site excludes the occupation of all adjacent sites.  With
this, the fermion creation operator becomes
$d_i^\dagger=c_i^\dagger {\cal P}_{<i>}$, where
\begin{equation}
{\cal P}_{<i>}= \prod_{j\hbox{ next to } i} (1-c^\dagger_j c_j)
\label{proj}
\end{equation}
is zero if any site next to $i$ is occupied. The hamiltonian is
defined in terms of the supercharge $Q=\sum_i d^\dagger_i$:
\begin{equation}
H=\{Q,Q^\dagger\}= \sum_{<ij>} d^\dagger_i d_j +  \sum_i {\cal P}_{<i>}.
\label{ham}
\end{equation}
The model is supersymmetric because $Q^2=(Q^\dagger)^2=0$, which
then implies that $[Q,H]=[Q^\dagger,H]=0$. The second term in the
hamiltonian combines a chemical potential and a repulsive
next-nearest-neighbor potential. The details of this term depend
on the lattice we choose.

A powerful analytical tool is the Witten index
\begin{equation}
W = {\rm Tr}[(-1)^F]
\end{equation}
with $F$ the operator counting the number of fermions in a given state. One
easily shows that $W=N_b-N_f$, where $N_b$ is the number of bosonic ground
states (those with an even number of fermions), and $N_f$ is the number of
fermionic ground states (those with an odd number of fermions) \cite{Witten}.
By this result, $|W|$ is a lower bound on the number of ground states. In analyzing
the supersymmetric model we also used techniques from cohomnology theory,
relying on a one-to-one correspondence between quantum ground states and
cohomology classes of the complex associated with the supercharge $Q$.

The papers \cite{fendley-2003-90,beccaria,fendley-2005-95,vaneerten,HHFS}
explored this model on a variety of lattices and revealed a number of remarkable
quantum phases. While some of the specifics of these phases require the fine-tuning
set by supersymmetry, many qualitative features are expected to survive in models
that are sufficiently close to the supersymmetric point.

On 1D lattices (quantum chains), the model turns out to be quantum
critical, with the critical behavior fully described by the first ($k=1$) unitary minimal
model of $N=2$ supersymmetric Conformal Field Theory (SCFT) \cite{fendley-2003-90}.

On generic 2D and 3D lattices, the supersymmetric lattice model displays a
phenomenon we call  {\bf superfrustration}. This term denotes a strong form
of quantum charge frustration, with the number of quantum ground states growing
exponentially with the volume of the system, implying  extensive ground state entropy
\cite{fendley-2005-95,vaneerten}.  For an elaborate account we refer the reader
to an earlier review \cite{proc}.

In this paper we focus on 2D lattices which exhibit yet another type of anomalous
behavior, with the number of quantum ground states growing exponentially with the
\emph{linear} dimensions of the system (sub-extensive ground state entropy). One
example is the 2D square lattice, where this property is established via a rigorous
theorem relating the number of quantum ground states to specific rhombus tilings
of the plane. A second example is the octagon-square lattice. In this latter case the
ground state structure is less involved, allowing us to delve a bit deeper and to
extend the analysis to the presence of defects. With some of the features observed
(torus degeneracies, presence of edge modes) being reminiscent of those of
topological phases of 2D matter, we tentatively refer to these phases as
`supertopological'.

\section{Square lattice}\label{sec2}
\begin{table}
\tbl{Witten Index for $M \times N$ square lattice.}
{\scriptsize
\begin{tabular}{r|rrrrrrrrrrrrrrrr}
 & 1 & 2 & 3 & 4 & 5 & 6 & 7 & 8 & 9 & 10 & 11 & 12 & 13 & 14 & 15\\
\hline
1  & 1 &  1 & 1 & 1 &  1 &  1 & 1 & 1 & 1 & 1  & 1 & 1 & 1 & 1 & 1\\
2  & 1 & -1 & 1 & 3 &  1 & -1 & 1 & 3 & 1 & -1 & 1 & 3 & 1 & -1& 1\\
3  & 1 &  1 & 4 & 1 &  1 &  4 & 1 & 1 & 4 & 1  & 1 & 4 & 1 & 1 & 4 \\
4  & 1 &  3 & 1 & 7 &  1 &  3 & 1 & 7 & 1 & 3  & 1 & 7 & 1 & 3 & 1 \\
5  & 1 &  1 & 1 & 1 & -9 &  1 & 1 & 1 & 1 & 11 & 1 & 1 & 1 & 1 & -9 \\
6  & 1 & -1 & 4 & 3 &  1 & 14 & 1 & 3 & 4 & -1 & 1 & 18 & 1 & -1& 4 \\
7  & 1 &  1 & 1 & 1 &  1 &  1 & 1 & 1 & 1 & 1  & 1 & 1 & 1  & -27 & 1\\
8  & 1 &  3 & 1 & 7 &  1 &  3 & 1 & 7 & 1 & 43 & 1 & 7 & 1  & 3 & 1\\
9  & 1 &  1 & 4 & 1 &  1 &  4 & 1 & 1 & 40& 1  & 1 & 4  & 1 & 1 & 4\\
10 & 1 & -1 & 1 & 3 & 11 & -1 & 1 & 43& 1 & 9  & 1 & 3  & 1 & 69 & 11\\
11 & 1 & 1  & 1 & 1 & 1  & 1  & 1 & 1 & 1 & 1  & 1 & 1  & 1 &  1 &  1\\
12 & 1 & 3  & 4 & 7 & 1  & 18 & 1 & 7 & 4 & 3  & 1 &166 & 1 &  3 & 4\\
\end{tabular}
}
\label{tab:square}
\end{table}

\begin{figure}[ht]%
\begin{center}
 \parbox{2.6in}{\epsfig{figure=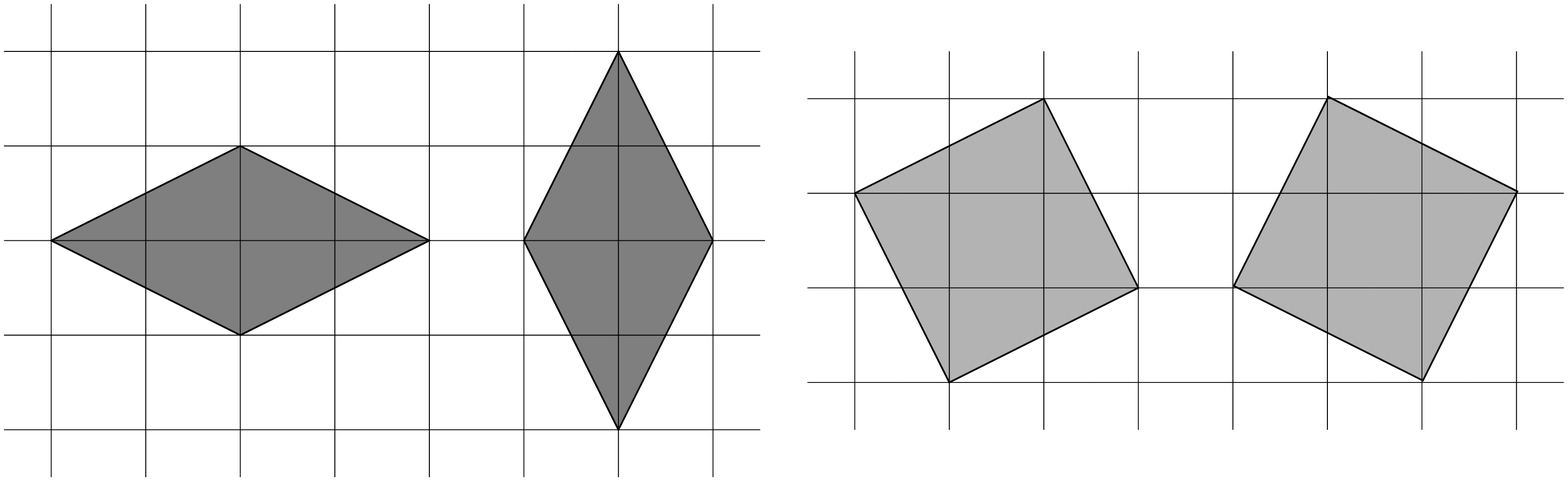,width=2.5in}
 }
 \hspace*{6pt}
 \parbox{1.1in}{\epsfig{figure=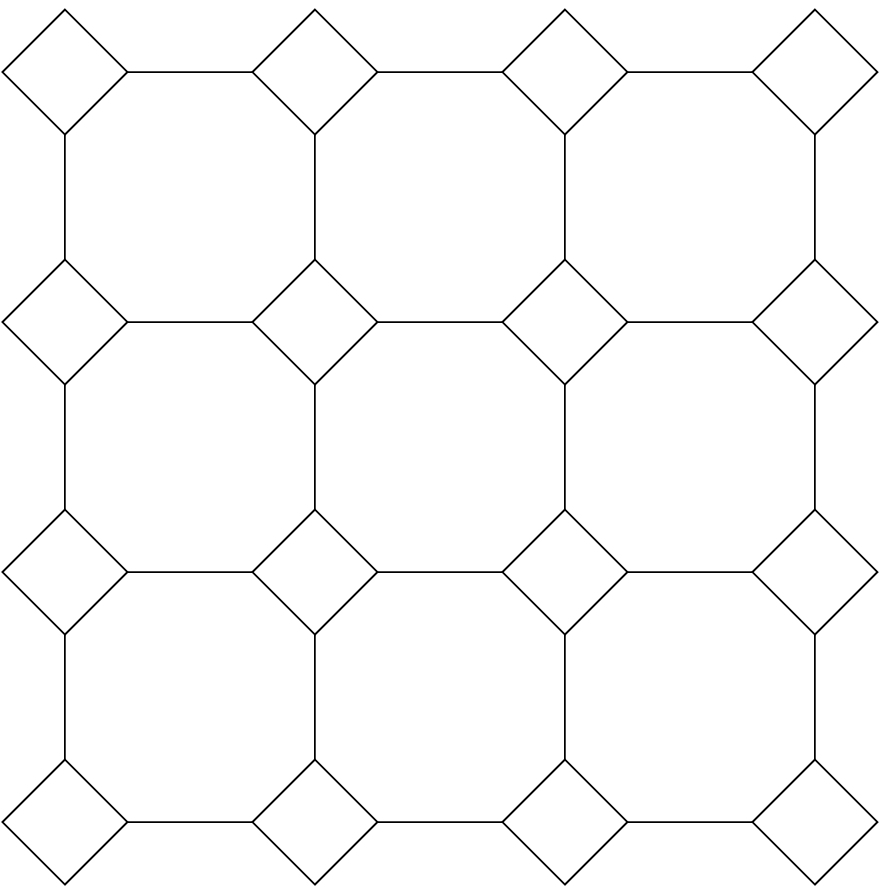,width=1in}
 }
 \caption{On the left the four rhombuses on the square lattice, on the right the
 octagon-square lattice of size $4\times4$.}
\label{fig}
\end{center}
\end{figure}

Where numerical studies of the Witten index \cite{fendley-2005-95,vaneerten}
showed superfrustration for a
wide range of 2D lattices, they revealed a very different behavior for the square
lattice wrapped around the torus (see Table~\ref{tab:square}). At first glance
one notices that the index does not grow exponentially with the system size as it
does for the superfrustrated systems. Inspired by the curiosities of this table and two
conjectures \cite{FSvE} on its structure, Jonsson \cite{Jonsson} proved a
general expression for the Witten index $W_{u,v}$ of the square lattice with
periodic boundary conditions given by the vectors $\vec{u}=(u_1,u_2)$ and
$\vec{v}=(v_1,v_2)$. He showed that $W_{u,v}$ is simply related to tilings constructed from
the rhombuses pictured in Fig.~\ref{fig}. Precisely, let $t_b$ ($t_f$) be the
number of ways of tiling the torus with these four rhombus types, so that there are an
even (odd) number of rhombuses.
\begin{theorem}[Jonsson, 2006]
The expression for the Witten index reads
\begin{equation}
W_{u,v} = N_b - N_f = t_b - t_f - (-1)^d \theta_d \theta_{d^*},
\end{equation}
where $d\equiv \hbox{gcd}(u_1-u_2,v_1-v_2)$,
$d^* \equiv \hbox{gcd}(u_1+u_2,v_1+v_2)$ and
\begin{equation}
\theta_d \equiv \left\{ \begin{array}{ll}
2 & \textrm{if $d=3k$, with $k$ integer}\\
-1 & \textrm{otherwise.}
\end{array} \right.
\end{equation}
\end{theorem}

A natural extension of Jonsson's theorem is to relate the {\em total} number of ground
states to rhombus tilings. Exploiting the one-to-one correspondence between ground
states and elements of the cohomology of the supercharge $Q$, we were able to prove
this relation explicitly when $\vec{u}=(m,-m)$ and $v_1+v_2=3p$ \cite{HHFS,HS}.
\begin{theorem}[Fendley, Huijse and Schoutens, 2009]
The total number of ground states reads
\begin{equation}
N_b + N_f = t_b + t_f+ \Delta,
\label{PF}
\end{equation}
where $|\Delta|=|\theta_d \theta_{d^*}|$. For $\vec{u}=(m,-m)$ and $v_1+v_2=3p$ we
find $\Delta=-(-1)^{(\theta_m+1)p} \theta_d \theta_{d^*}$.
\end{theorem}
In this correspondence, the number of fermions in a given ground state matches
the number of tiles in the corresponding tiling.
Although the proof is restricted to a certain set of periodicities, there is strong
evidence that the theorem holds for general $\vec{u}$ and $\vec{v}$. We explicitly
checked the result for a variety of small systems.

The relation between tilings and ground states implies that for large
enough systems ground states exist at all rational fillings (particles per site)
$F/N$ between $1/5$ and $1/4$ (see also \cite{Jo2} for an alternative proof).
One can show that for $\vec{v}=(n,n)$ and $m=3p$, $n=3q$, the number of
tilings grows as $4^{p+q}/\sqrt{pq}$, thus establishing sub-extensive
ground state entropy  \cite{Jo3}.

For free boundary conditions in either one or both of the diagonal directions along the
square lattice ($(m,-m)$ and $(n,n)$) the number of ground states reduces dramatically
\cite{B-M,HHFS}. One finds that it is either one or zero, except for the
cylindrical case periodic in the $(m,-m)$-direction with $m=3p$ and $n=3q+2$ or
$n=3q+3$. In that case the number of ground states is $4^{(q+1)}$.

With the above results in place, the ground state counting problem for the square
lattice has been completely settled. Further pressing questions concern the nature
of these ground states and of the excited state spectrum. Some progress on these
matters was provided in \cite{HHFS}, where we presented numerical results strongly
indicating the presence of critical modes in ladder versions of the 2D lattice. We
then argued that the full 2D lattice with (diagonal) open boundary conditions
supports edge modes described by $N=2$ superconformal field theory.

While the physical understanding of the quantum phase on the
square lattice remains far from complete, one is led to a picture
where the ground state corresponding to a given tiling has
fermions that are confined to the area set by an individual tile,
but quantum fluctuating within that space. For the particular case
of a ladder with periodicity vectors $(1,2)$ and $(L,0)$, closed
form expressions for the ground states at filling 1/4 confirm this
picture; for the more general case such explicit expressions are
not available. The tiling based physical picture of the ground
state wavefunctions is reminiscent of electrons in a filled
magnetic Landau level, each of them effectively occupying an area
set by the strength of the magnetic field. Critical edge modes
naturally fit  into a picture of this sort.

\section{Octagon-square lattice}\label{sec3}

A second 2D lattice where the supersymmetric model displays
sub-extensive ground state entropy is the octagon-square lattice
(Fig.~\ref{fig}). The growth behavior of the numbers of ground
states on the plane, cylinder and torus is similar to that of the
square lattice. A big difference, however, is that here all ground
states reside at $1/4$ filling. This hugely simplifies the
computation of the degeneracies. For the plane we find that the
ground state is unique. For the cylinder with $M \times L$ square
plaquettes, where $M$ is the number of square plaquettes along the
periodic, horizontal direction and $L$ along the open, vertical
direction, the number of ground states is $2^L$. Finally, for the
torus the number of ground states is $2^M+2^L-1$
\cite{fendley-2005-95}.

There is again a relatively simple physical picture which we propose as
a basis for further analysis of physical properties. This picture reflects the
systematics uncovered by the analysis of the associated cohomology
problem as well as results for small system sizes. The basic building block
of the many-body ground states is the 1-fermion ground state on an isolated
square plaquette. The unique many-body ground state on the plane essentially
has individual fermions occupying this
lowest 1-plaquette orbital, again allowing the analogy with a filled Landau level.
Closing boundaries leads to the possibility that electrons on horizontal or vertical
rows of plaquettes `shift' into a second 1-fermion state, this way building up the total
of $2^M+2^L-1$ ground states. This picture can be further substantiated by
allowing defects, which we bring in by adding diagonal connections in individual
plaquettes.

Among the key issues that are presently on the agenda for further study are:
the existence of energy gaps, the presence of bulk or edge critical modes, and
interactions and braiding properties of defects. We are confident that the constraints
set by the supersymmetry, which have been instrumental in the progress made so
far, will allow further progress in the analysis of the remarkable `supertopological'
phases on the square and octagon-square lattices.

\bibliographystyle{ws-procs975x65}
\bibliography{ws-pro-sample}

\end{document}